# The Submillimeter Array – Antennas and Receivers

Raymond Blundell
*rblundell@cfa.harvard.edu*
*Harvard-Smithsonian Center for Astrophysics*
*60 Garden Street, Cambridge, MA 02138*

**Abstract**

The Submillimeter Array (SMA) was conceived at the Smithsonian Astrophysical Observatory in 1984 as a six element interferometer to operate in the major atmospheric windows from about 200 to 900 GHz. In 1996, the Academica Sinica Institute of Astronomy and Astrophysics of Taiwan joined the project and agreed to provide additional hardware to expand the interferometer to eight elements. All eight antennas are now operating at the observatory site on Mauna Kea, and astronomical observations have been made in the 230, 345, and 650 GHz bands.

The SMA antennas have a diameter of 6 m, a surface accuracy of better than 25 μm rms, and can be reconfigured to provide spatial resolutions down to about 0.5" at 200 GHz and, eventually, 0.1" at 850 GHz. Coupling to the receiver package within each antenna is achieved via a beam waveguide, in a bent Nasmyth configuration, comprised of a flat tertiary mirror and two ellipsoidal mirrors that form a secondary pupil used for receiver calibration. An additional fixed mirror and a rotating wire grid polarizer are then used for receiver selection. Each antenna houses a single cryostat, with an integrated cryocooler capable of cooling up to eight receivers to 4 K. In the current configuration only three receiver bands are available: 175 – 255 GHz, 250 – 350 GHz, and 600 – 720 GHz, and simultaneous operation of the 650 GHz receiver with either of the lower frequency receivers is possible. Eventually dual polarization will be available from 325 – 350 GHz, and dual frequency operation will be possible, pairing either of the lower frequency receivers with any of the high frequency units: 325 – 425 GHz, 425 – 510 GHz, 600 – 720 GHz, and 800 – 900 GHz.

Each receiver currently uses a single superconductor-insulator-superconductor junction as the mixing element, and has first stage intermediate frequency amplification at 4 K with an instantaneous bandwidth of 2.5 GHz, centered at 5 GHz. The mixers are of a fixed-tuned waveguide design, are inherently broad band, typically 80 – 100 GHz, and provide for relatively low receiver noise, typically 5 – 10 hv/k for the low frequency bands, and 10 – 15 hv/k for the 650 GHz receivers. Local oscillator power to each receiver is provided by a mechanically tunable Gunn oscillator followed by the appropriate diode frequency multiplier or multiplier combination.

**Introduction**

It has long been recognized that the wavelength range 0.3 to 1.3 mm observable with ground-based telescopes offers unique opportunities in the study of cool (10 – 100 K) dust and gas clouds in the Milky Way and other galaxies. By the mid 1980's the California Institute of Technology 10 m diameter telescope (the Caltech Submillimeter Observatory), and the James Clerk Maxwell Telescope (JCMT), a 15 m diameter instrument of the United Kingdom, the Netherlands, and Canada were under construction specifically for observations in the submillimeter. The 10 m diameter Submillimeter Telescope, a collaborative effort between the University of Arizona and the Max Planck Institut für Radioastronomie, was also under construction on Mount Graham in Arizona. In addition, the 15 m diameter Swedish European Submillimetre Telescope (SEST), designed predominantly to access the southern sky at millimeter wavelengths, was also under construction at the European Southern Observatory facility of La Silla which borders the southern extremity of the Atacama Desert in northern Chile. At their shortest wavelengths these instruments had an angular resolution of approximately 6" – 15".

Also during the mid 1980's, the pioneering millimeter wavelength interferometers at Hat Creek (University of California, Berkeley) and at Owens Valley (California Institute of Technology) offered spatial resolutions of somewhat less than 5". Two other interferometers, at the Plateau de Bure in the French Alps and at Nobeyama were being developed by the Institut de Radio Astronomie Millimétrique and by the

National Astronomical Observatory of Japan, respectively. Both of these instruments were designed to offer higher spatial resolution ~ 1". Furthermore, having demonstrated high resolution imaging through interferometry at millimeter wavelengths, and with the anticipated success of the CSO and JCMT, a natural step forward was to propose to design and construct a submillimeter interferometer. Such an instrument could be used to provide additional high resolution information about, for example, the solar system, star formation, astrochemistry, the structure of galaxies, and the energetics of quasars and active galactic nuclei.

In 1984, following a request by the Director of the Harvard-Smithsonian Center for Astrophysics, Dr. Irwin I. Shapiro, the Submillimeter Telescope Committee, headed by Dr. James M. Moran, proposed the design and construction of a Submillimeter Wavelength Telescope Array [1]. They envisioned an array of six 6 m diameter antennas, situated at a dry high altitude site, configured to offer sub arc-second resolution in the wavelength range 0.3 to 1.3 mm. The committee also recognized that the investment in receiver technology would be both substantial and crucial to the ultimate performance of the array, and recommended that the development of a receiver laboratory be given the highest priority. Start-up funding was received in 1989 to set up such a lab, and SAO's Major Scientific Instrumentation fund, set up in 1992, has been used to fund the design and construction of the array. Finally, in 1996, the Academica Sinica Institute of Astronomy and Astrophysics of Taiwan joined the SMA project and agreed to provide an additional two antennas and associated hardware to enable eight element operation.

From experience at existing millimeter and submillimeter telescope sites it was evident that a high altitude plateau of reasonable extent (>500 m) would be required to site the SMA. Furthermore, in order to keep construction costs at a reasonable level, a mainland site with existing infrastructure, such as access, electricity supply, local industry, etc. would be preferable. In 1984, Mount Graham was believed to be the best of the candidate mainland sites; Jelm Mountain in Wyoming was quickly eliminated due to its high latitude; and the Aquarius Plateau in Utah, Chalk Mountain in Colorado, and the White Mountain in California were also identified as worthy of further study. Outside the continental US, the South Pole was known to offer an extraordinarily dry environment, but was eliminated due primarily to extreme logistical problems, and the summit of Mauna Kea in Hawaii was believed to be a good candidate site. Subsequently, a number of large-area high-altitude locations in northern Chile were also identified as candidate sites [2]. Finally, based primarily on available site-testing data and local infrastructure, and the proximity of the CSO and the JCMT, Mauna Kea was selected as the site for the SMA in 1992.

Having selected Mauna Kea as the SMA site, it was immediately obvious that, unlike existing millimeter wavelength interferometers, which make use of rail (Nobeyama, Owens Valley, Plateau de Bure) or paved road (Hat Creek) for array reconfiguration, the SMA antennas would have to be moved over more difficult terrain. Any reconfiguring of the array to change angular resolution consequently excluded the use of protective radomes, so that the SMA antennas would be unprotected from the environment. This, coupled with the fact that abrasive volcanic particles are often blown about the summit and severe icing conditions can occur in periods of high wind, effectively ruled out the use of aluminized carbon fiber panels for the antenna reflector surface. Finally, also in 1992, after a letting number of design study contracts to industry, SAO became prime contractor for the design and construction of the SMA antennas.

**Antennas**

In order to proceed with the SMA antenna design, the Mauna Kea specific environmental conditions, set out in Table 1, were developed [3]. Referring to the table, we should note that the nominal operating environment was specified to exceed the $90^{th}$ percentile conditions, so observing under the so-called 'worst case' conditions would be unusual. While the general specifications for the SMA antennas are given with reference to Table 2 below, a number of other factors not listed as specifications were deemed critical. For example, the desire to reconfigure the array in a day implied moving a single antenna to a new location over unpaved road with up to 15% grade in 1 hour. Also, the requirement to maintain a physically stable environment with respect to gravity, for the cryogenically cooled receivers, effectively ruled out secondary focus operation. In other words, antenna designs of only the Nasmyth or Coudé configuration were considered.

In addition to the general specifications laid out in Table 2, there was also a requirement that the antenna remain sufficiently stable to enable blind operation, i.e. without pointing checks, under the nominal operating environment for periods of at least 1 hour. For the average operating environment in which thermal effects are only half the nominal, and for wind speeds < 10 ms$^{-1}$, this implied a pointing accuracy of better than 2" rms.  A phase stability requirement, and a minimum observing interval of 20 minutes

Nominal operating environment
- Wind: 14 ms$^{-1}$ average, 20 ms$^{-1}$ peak
- Air temperature: -15 to +25 C with 10 C diurnal fluctuations

Worst case operating environment (including antenna transport)
- Wind: 25 ms$^{-1}$ average
- Air temperature: -20 to +30 C day and night operations

Survival environment
- Wind: 60 ms$^{-1}$ average
- Rain: 0.1 mhr$^{-1}$ for 1 hour
- Snow: 0.5 m
- Ice: 10 mm on flat surfaces, 75mm on tubular structures

**Table 1: Environmental considerations for antennas placed on Mauna Kea.**

between calibrations, resulted in the additional requirement that the antenna cabin housing the receivers and other sensitive electronic equipment be temperature controlled to ±1 C.  Other specifications, such as antenna leveling and the repeatability of antenna placement after transport were not defined at that time.  However, the azimuth axis verticality was specified to be better than 15".

| | |
|---|---|
| Antenna configuration | Alt – Az |
| Operating range | Azimuth ±270° |
| | Elevation 5 – 90° |
| Sun exclusion zone | None |
| Intersection of axes | < 0.10 mm |
| Non-perpendicularity of axes | < 10" |
| Azimuth bearing run-out | < 30 μm axial and radial |
| Non repeatable run-out | < 2 μm both axes |
| Azimuth bearing wobble | < 3" |
| Non repeatable wobble | < 0.2" |
| Slew rate | > 4°s$^{-1}$ azimuth, > 2°s$^{-1}$ elevation |
| Tracking accuracy | 1.3" rms per axis |
| Tracking resolution | 0.3" rms per axis |
| Absolute pointing accuracy | 2.8" rms per axis |
| Antenna cabin (receiver enclosure) | 2.6 m x 2.6 m x 2 m high |
| Nominal reflector diameter | 6 m |
| Antenna surface | Aluminum panels |
| Antenna efficiency | > 50% at 350 μm |
| Secondary mirror translation | X and Y ±7 mm (50 μm resolution) |
| | 50 mm travel in Z (2 μm resolution) |
| Non-chopping position | Centered to ±3" to give ±0.3" on sky |
| Chopping amplitude | ±24' to yield ±2.5' on sky |
| Chopping rate | DC to 10 Hz |

**Table 2: General specifications for the SMA antenna.**

Following the specifications laid out in Table 2, the SMA mechanical engineering staff developed an antenna design based on a bent Nasmyth configuration. Referring to Figure 1 (Left), the design of the antenna pedestal was driven primarily by the stiffness required to maintain pointing and phase accuracy for the expected wind loads, and a detailed thermal analysis determined that an all steel structure could maintain the required performance as long as it is properly isolated from the external environment. To this end, the bulk of the structure is housed inside the receiver cabin which is temperature controlled to ±1 C, and the base, below the azimuth bearing, is thermally insulated from its surroundings.

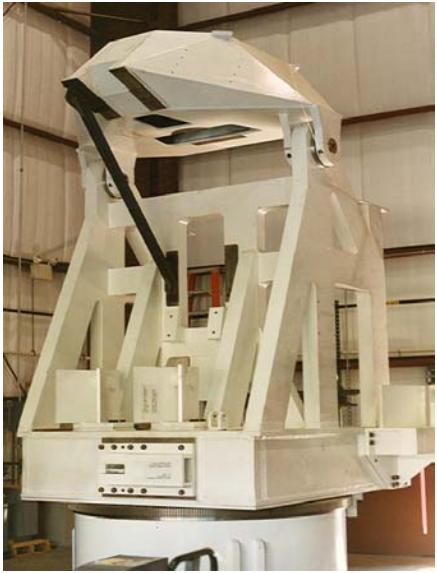

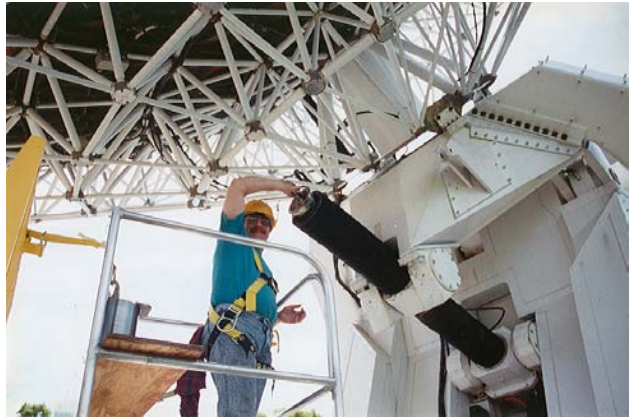

**Figure 1:** Left, the SMA antenna mount, made of 130 mm thick steel plate to meet the antenna stiffness requirements for pointing. Above, detail of the interface between the antenna pedestal and the reflector assembly. Note the ball-screw drive and the four flex-pivots used to connect the carbon fiber central hub of the reflector assembly to the steel weldment that connects to the elevation axis and screw drive.

Having eliminated the use of aluminum coated carbon fiber reflector panels, despite their inherently better temperature stability and light weight, the SMA antenna reflector assembly was designed to incorporate machined aluminum panels. Furthermore, as thermal considerations excluded the use of panels no larger than 1 m in extent, a 4-ring structure was chosen with 12 panels in the innermost ring and 24 in rings 2, 3, and 4. The panels are supported above the steel nodes of a back-up structure made up of a truss-work of carbon fiber tubes. Apart from the inner ring of panels which each has 3 support points, a redundant four point panel support scheme was chosen. In this way, individual panels can be deformed in-situ to correct for any large scale manufacturing defects. Finally, a quadrupod, made from carbon fiber in order to meet focus and pointing stability requirements, was included to support the chopping secondary mirror assembly.

Referring to Table 3, in which cross polarization effects and other minor losses have been neglected, we split the signal losses due to an imperfect antenna into two main categories. The first depends strongly on the particular type of antenna design chosen and the second, apart from the blockage component, should be considered fixed and unrelated to the particular choice of antenna design. In other words, in order to meet the overall antenna efficiency specification of better than 50% at 350 μm wavelength, the sum of the antenna dependent efficiencies should exceed 0.65. Given that the sum of the projected antenna efficiencies, 0.57, is discrepant, it is useful to examine the sources of the largest errors: focus, phase, pointing, and Ruze, in more detail. While it may be possible to correct for some focus errors, such as deflections due to gravity or global ambient temperature changes, deflections due to wind loads and temperature gradients present more of a problem. These errors were calculated, for the nominal operating wind load of 14 ms$^{-1}$ and worst case thermal gradients, via structural and thermal models, and were transformed to an error in separation along the antenna bore-sight between the focus of the secondary mirror and the focus of the best-fit deflected parabola of the primary. The efficiency due to focus errors [4] at a wavelength of 350 μm was then calculated using

$$\eta_f = \exp{-6.5(\varepsilon_f)^2}$$

where the focus error, $\varepsilon_f$, is expressed in millimeters [3]. Path length errors, or phase errors, are somewhat related to focus errors, and are especially relevant to interferometer operation in which the difference in variation of path length between the astronomical source and the receiver in each antenna results in a de-correlation of the interferometer signal and a consequent loss of gain. Fortunately, changes in path lengths that affect all antennas equally and simultaneously do not produce a resultant error, and almost all of the path length errors are therefore due to variations in the local environmental conditions at each antenna. The efficiency due to path length errors was calculated, assuming that three of the six antennas have path length errors while the path length to the other antennas remains constant [5], using

$$\eta_{pl} = \exp{-650(\varepsilon_{pl})^2}$$

where the path length error, $\varepsilon_{pl}$, is expressed in millimeters [3]. For single dish operation path length errors would not be an issue, $\eta_{pl} = 1$, so the projected antenna efficiency at 350 μm would be 50%. Pointing errors are defined as the combination of the residual error of the antenna pointing model and quasi-static errors resulting from a steady wind and thermal distortions of the antenna mount and reflector assembly. Short-term pointing fluctuations due to wind gusts and servo errors fall under the category of pointing jitter. The efficiency resulting from pointing errors [5] was calculated using

$$\eta_p = \exp{-0.017(\varepsilon_p)^2}$$

where the pointing error, $\varepsilon_p$, is expressed in arc-seconds [3]. Finally, we consider the loss in efficiency due to phase front errors arising from an imperfect reflector surface. Referring to Table 4 we note that, largely as a result of improved thermal performance, the use of carbon fiber reflector panels would reduce the sum of the panel errors to about 8 μm rss and result in a reflector efficiency of 87%. However, the SMA baseline antenna design excluded the use of carbon fiber panels on the grounds of durability given the harsh environment of Mauna Kea, and the projected reflector efficiency is somewhat higher.

| | |
|---|---|
| Focus errors | 0.92 |
| Ohmic losses | 0.99 |
| Phase errors | 0.89 |
| Pointing errors | 0.87 |
| Pointing jitter | 0.97 |
| Surface errors | 0.84 |
| Sum of antenna dependent errors | 0.57 |
| Blockage | 0.93 |
| Alignment errors | 0.95 |
| Illumination | 0.87 |
| Sum of other 'fixed' errors | 0.77 |
| Total | 0.44 |

**Table 3:** Efficiency error budget for the SMA antennas, designed to the general specifications given in Table 1, under the nominal operating environment.

We should also note that, under average operating conditions in which thermal effects are only half the nominal, and for wind speeds < 10 ms$^{-1}$, this design should result in an antenna efficiency greater than 50% at a wavelength of 350 μm. Furthermore, since it is extremely unlikely that the worst case thermal conditions will occur in winds approaching the nominal 14 ms$^{-1}$, it is likely that this design will provide an overall antenna efficiency greater than 50% at a wavelength of 350 μm under the nominal operating conditions.

| | |
|---|---|
| Panel accuracy (manufacturing) | 5 μm rms |
| Panel setting errors | 5 μm rms |
| Other panel errors (mainly thermal) | 7 μm rms |
| Sum of panel errors | 10 μm rss |
| Backup structure (BUS) | Carbon fiber with steel nodes |
| BUS errors (mainly thermal + gravity) | 6 μm rss |
| Sum of reflector errors | 12 μm rss |
| Reflector efficiency (surface errors) | 84% |

**Table 4: Antenna surface error budget for the nominal operating environment.**

The requirement to have a relatively large receiver enclosure that was fixed in elevation resulted in the use of the ball-screw arrangement for the antenna elevation drive. In azimuth, a more standard drive system, consisting of two motors driven in opposition against a spur gear cut directly on the azimuth bearing of the antenna, was selected. In both cases high torque, 1,100 Nm, brushless servo motors were chosen, in part to fulfill the desire for fast antenna position switching. In order to provide the required controlled antenna motion and to easily accommodate the variability of the elevation gear ratio a digital servo control drive system was implemented. In simple terms, the antenna servo control consists of three nested feedback loops. The innermost, the motor commutation loop, consists of the brushless servo motor, position resolver, and servo amplifier, and includes a hardware safety function to limit the maximum velocity the motor can attain. The second feedback loop is the velocity loop in which a tachometer is used to provide the servo control board (SCB), designed and programmed by receiver lab staff, with a direct feedback of the motor's speed. The outermost loop, the position loop, incorporates a 23 bit high resolution encoder and issues the SCB with a velocity command, based on the desired position, via the real-time antenna control computer (LynxOS) which communicates with a central computer in the SMA control building. New velocity requests are computed and transmitted to the SCB at 100 Hz, but the effective position loop bandwidth is about 5 Hz in both axes. The elevation velocity loop uses PID gains, while the azimuth velocity loop uses only PI terms. Finally, a Palm Pilot can be used to control the antenna via the SCB for maintenance and system diagnostics.

The quality of the servo control system as a whole is best assessed through the tracking and slewing performance of the antennas measured at the observatory site on Mauna Kea. Under calm conditions we typically achieve sidereal-rate tracking errors of ~ 0.3" rms on both axes on all six SAO built antennas. To demonstrate performance we plot the response of antenna 4 tracking and switching between a source and calibrator in strong winds in Figure 2. The tracking specification of < 1.3" rms is clearly met for both axes, suggesting that losses in antenna efficiency as a result of antenna drive or tracking errors are likely to be small. However, the data presented are based on antenna encoder readings and do not include errors in absolute pointing, or errors due to misplacement of the chopping secondary mirror. Errors in the absolute pointing of the antenna are removed with the help of an optical guide-scope and a multi-parameter mount model in the usual way and residual pointing errors of 2" rms are typical for the SMA antennas. In addition to an optical pointing model, models of the radio pointing of each SMA antenna are made on a regular basis and radio to optical offsets are incorporated into each antenna pointing model. Referring to Figure 2, we note that the slew rate specifications of 4°s$^{-1}$ in azimuth and 2°s$^{-1}$ elevation have been met, and a duty cycle of better than 50% is achieved for on-source times as short as 10 s. However, with a source acquisition time of ~ 3 s in azimuth, the observing efficiency degrades significantly for shorter timescales.

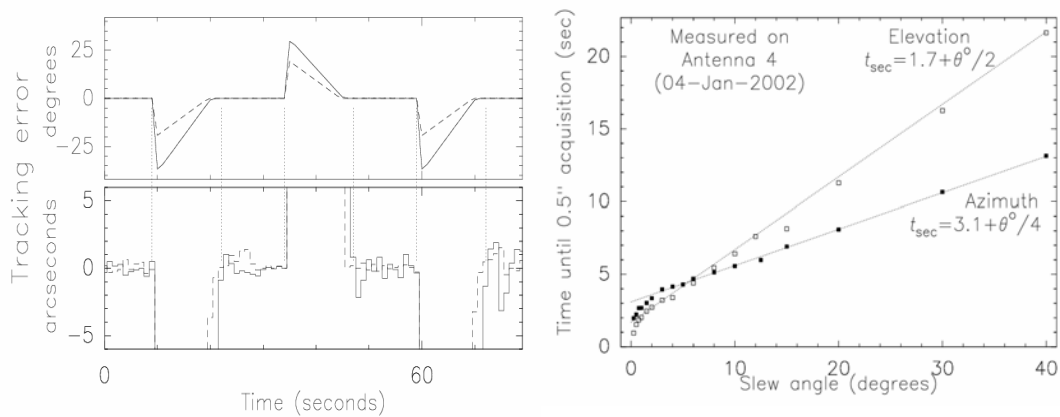

**Figure 2:** Left, demonstration of the tracking and slewing performance of the SMA antennas in high winds. During the time the antenna is either 'on-source' or tracking a calibrator, the specification of < 1.3" rms is easily met. Right, fast switching of the antennas is ultimately limited by the source acquisition time of 3 s in azimuth.

During assembly, each SMA reflector surface was first set to ~ 60 μm rms using a mechanical swing template equipped with capacitive sensors. The surface of the first of the SMA reflectors was improved using a near-field holographic technique employing a transmitter at 94 GHz and two purpose-built room-temperature Schottky diode mixer receivers. One of these, the reference receiver, was mounted behind the secondary mirror, the other was mounted on the elevation axis and was coupled to the SMA antenna via a flat tertiary mirror. As soon as two SMA antennas, equipped with working low-noise SIS receivers designed for astronomical use, became available the holographic technique was extended to 232.4 GHz, and surface adjustments are now based on measurements that include the complete set of receiver and antenna optics. Simply put, the complex beam pattern of the antenna under test is measured using on-the-fly mapping with a second antenna in the array providing the phase reference. The resultant beam pattern is Fourier-transformed to produce amplitude and phase distributions on the aperture of the antenna under test. This aperture phase is then corrected for systematic errors due to the fact that the measurements are made in the near-field, leaving behind phase errors attributable to surface deviations from a best-fit paraboloid. This gives a surface error map from which a list of individual panel adjustments can be derived. Figure 3 shows a plot of the surface of antenna 4 as originally set using the swing template, and a plot of the surface after three rounds of surface adjustments based on holographic data taken at 232.4 GHz. All of the SMA antennas have now been set to better than 25 μm rms using this technique. At the present time only one antenna has been set to better than the specification of 12 μm rms, and verification of performance at different elevations has yet to be demonstrated through celestial holography.

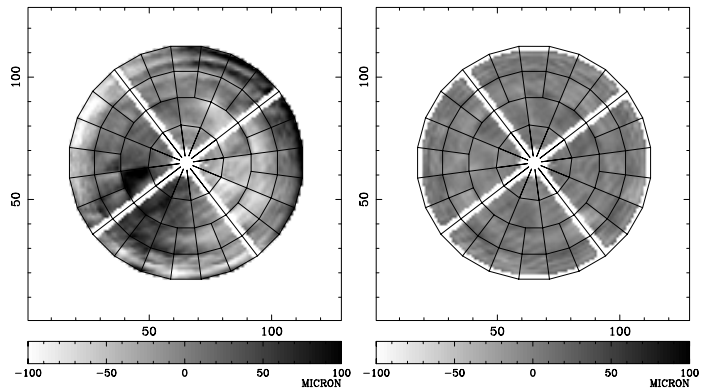

**Figure 3: The SMA antenna reflectors are typically set to 60 μm rms using the mechanical swing template. A near-field holographic technique, using a phase-locked transmitter at 232.4 GHz, is then used to progressively improve the surface. Two to four iterations are usually required to reach the design specification of 12 μm rms.**

The preliminary design for the SMA beam waveguide to relay the signal beam from the subreflector to the receivers excluded the use of curved mirrors primarily for ease of optical alignment and freedom from mirror-induced cross-polarization [3]. However, the all flat design introduced problems relating to frequency dependence and beam truncation. For example, a large tertiary mirror > 500 mm in extent would have been required in order to keep truncation losses at an acceptable level. In order to avoid these problems two ellipsoidal mirrors, mirrors M4 and M5, were introduced in the horizontal plane of a bent Nasmyth configuration after the tertiary mirror, and an additional flat mirror, M6, was introduced to direct the signal beam downwards and into the receiver package. Coupled with a lens in front of each mixer, mirrors M5 and M4 are used to produce a frequency independent image of the feed horn aperture at the subreflector. Furthermore, referring to Figure 4 (left), a compact image of the feed is produced between these mirrors providing a good location for receiver gain calibration and for insertion of the quarter wave plates required for polarization measurements.

Besides providing efficient coupling to the receivers, the antenna must also provide a phase stable environment for the entire signal path within the antenna receiver cabin including the beam waveguide, receiver optics, reference frequency distribution, and intermediate frequency transmission. For example, a 1 C temperature change in a 2 m long optical train, supported by a structure made predominantly from steel, would result in a 1 radian phase change at the highest operating frequency of the SMA. For this reason, and to ensure adequate thermal stability of the antenna mount, a temperature stability requirement of ±1 C was therefore imposed on the receiver enclosure. In order to achieve this stability we developed an air handler; basically a blower and two dampers that mix inside and outside air in the proper proportions to obtain the desired temperature. One damper controls the flow of cool outside air into the receiver cabin, the other controls the amount of warm air, heated by passing through the electronics rack, reentering the cabin. In the event that the electronics rack does not dissipate enough heat to maintain the desired temperature, a heater located in the ductwork can be energized. This system can maintain the cabin air temperature to ±0.5 C for indefinite periods under a variety of atmospheric conditions. Furthermore, unlike commercial air conditioners no refrigerant material is required. A slight overpressure is maintained inside the cabin to keep dusty air out, outside air entering the cabin via the air handler is filtered and free of dust, and waste heat from the electronics rack is put to good use.

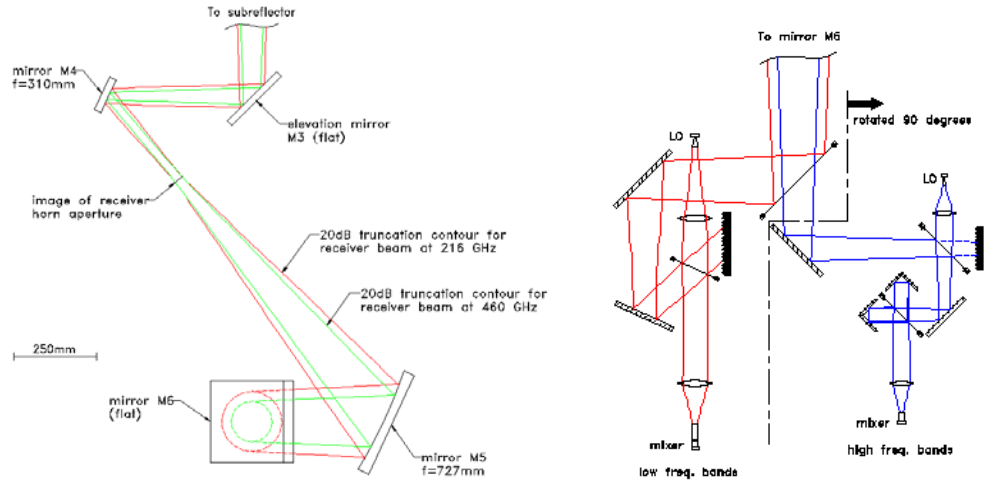

**Figure 4:** Left, two ellipsoidal mirrors are used in a bent Nasmyth configuration to provide a convenient location for receiver gain calibration and wave plates used for polarization measurements. Right, schematic of the optical arrangement used to couple multiple receivers to the SMA antenna. By a simple rotation of either the central mirror or wire grid assembly, any one of the four high frequency receivers can be used with either of the low frequency receivers, and dual polarization is possible in the frequency range 325 – 350 GHz.

### Receivers

Even though sensitivity, beam size, and angular resolution estimates for the array had been presented for 230, 345, 460, 690, and 850 GHz operation, as originally proposed [1] the SMA was to be equipped with heterodyne receivers operating in only three frequency bands. However, by 1994, it was clear that more receivers would be required to meet both scientific and technical requirements [4]. Referring to Table 5, the lowest frequency of operation was set by the need to be able to operate reliably and make useful astronomical observations in periods of average weather conditions, and for calibration and testing of the instrument. The highest frequency was set by the need to have reasonable atmospheric transmission during periods of good weather. The intermediate receiver bands were set primarily to ensure frequency coverage throughout the major atmospheric windows, and a dual polarization capability was included in order to permit efficient polarization measurements of the dust continuum in the frequency range of maximum sensitivity of the array, 330 – 350 GHz. Finally, in order to provide accurate phase calibration during the highest frequency observations, dual frequency operation, in which a low frequency receiver is paired with a high frequency receiver, was required.

| Receiver | Frequency (GHz) | 1984 Noise (K) | 1992 Noise (K) |
|---|---|---|---|
| 230 | 175 – 255 | 300 | 60 |
| 345 | 250 – 350 | 900 | 150 |
| 400 | 325 – 425 | | |
| 460 | 425 – 510 | 2000 | 200 |
| 650 | 600 – 720 | 5000 | |
| 850 | 800 – 900 | 8000 | 750 |

**Table 5:** The current SMA receiver plan calls for single pixel receivers to operate throughout the bulk of the major atmospheric windows from about 200 to 900 GHz. The first column gives the receiver designation, the second gives the sky frequency coverage, and the third and fourth columns respectively give the best measured SSB receiver noise at the time the SMA was proposed [1], and the projected SSB receiver noise performance figures from the SMA design study of 1992 [3].

Due to the large number of receivers the coupling optics were kept as simple as possible and the optical scheme, shown in Figure 4 (right), was developed to couple to the multiple receivers housed in a single cryostat. Referring to the figure, the lowest frequency receivers are coupled to M6 via the rotating wire grid assembly and local oscillator (LO) injection is achieved via a simple mesh coupler. As LO power is abundant at these frequencies, signal losses are kept below 1%. In contrast, the polarization of the high frequency receivers is orthogonal to the lowest frequency units, and the selection of the high frequency receivers is achieved via a rotation of the central mirror assembly. Since available LO power for the highest-frequency bands is more limited, LO power is coupled to the mixer using a Martin-Puplett interferometer. It is worth noting that because of the relatively wide IF bandwidth to IF center frequency of the SMA, 2 GHz centered at 5 GHz, the performance of the mesh coupler is actually superior to that of the Martin-Puplett when sufficient LO power is available.

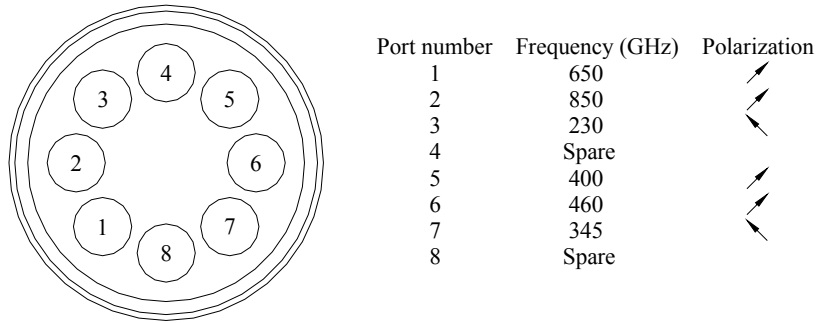

**Figure 5: Top view of the SMA cryostat, showing the relative position and polarization of the various receivers.**

Referring to Figure 5, the receivers in ports 3 and 7, the 230 and 345 GHz units, have radial polarization and are coupled to M6 by reflection off the rotating central wire grid assembly. The receivers in ports 1 and 5, the 650 and 400 GHz units, also have radial polarization, which is orthogonal to that of the 230 and 345 GHz units, and are coupled to M6 by reflection off the central rotating mirror assembly immediately beneath the wire grid assembly. Receivers to cover the 850 and 460 GHz bands will eventually be installed into ports 2 and 6, will couple to the antenna by reflection off the central rotating mirror assembly, and will have the same polarization as the 650 and 400 GHz receivers so that they can be used with either of the lowest frequency receivers.

The desire to operate with both low frequency and high frequency receivers simultaneously, either for phase calibration or for scientific purposes, imposes stringent requirements on receiver alignment. Referring to Table 2, the pointing requirement for efficient antenna operation was set at 2.8" rms per axis at 850 GHz. If we assume that, under dual frequency operation, each antenna is pointed with the aid of a pointing model, derived for the high frequency receiver in use, the low and high frequency receivers must be co-aligned to within about 6" in order to maintain a similar pointing loss for both receivers. With this in mind, implementation of the optical coupling of each mixer to the antenna via M6 relies heavily on a robust design concept, accurate machining, and sound metrology. After fabrication and careful alignment of each set of receiver optics, a near-field vector scanning technique, pioneered at SAO [6], is used to determine beam parameters from which minor mirror adjustments can be made to co-align each receiver. Ideally, the near-field scan should be made at either an image of the telescope focus (where pointing relates to lateral offset), or an image of the telescope pupil (where pointing relates to wave-front tilt.) In the SMA optics design a lateral offset of the beam results in a pointing offset of about 2" per mm in the plane of the image of the telescope focus, located about half way between the cryostat vacuum windows and M6.

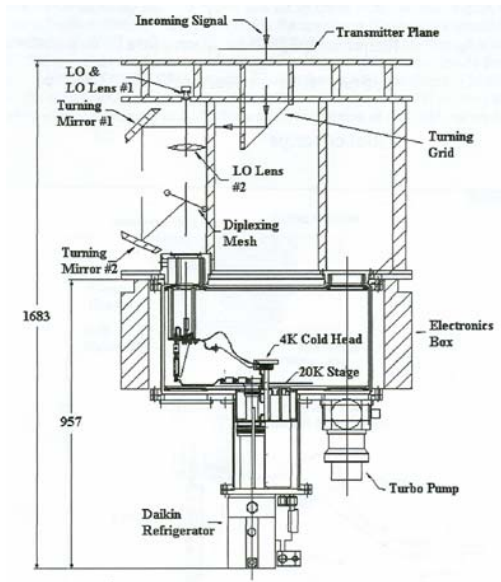
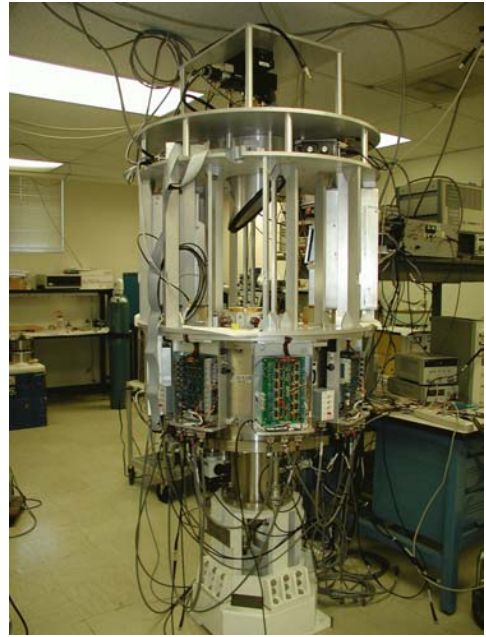

**Figure 6:** Left, schematic of the SMA receiver system showing the optics, receiver insert, and cryostat assemblies. Right, photograph of a complete SMA receiver system showing the cryostat surrounded by the receiver control electronics, the optics cage supporting the various sets of receiver coupling optics, and the central wire grid assembly used to select between the lower frequency receivers. The near-field scanning unit is also visible in place of M6 at the center and above the complete receiver system.

Unfortunately, due to mechanical constraints, near field scans cannot be made at this image of the telescope focus. Instead, scans are made in a plane just beneath M6. Nevertheless, fitting the data measured on a 71x71 grid with a 1 mm spacing generally results in co-alignment of the receiver beams to within 5" on the sky after a few iterations.

Since dual frequency operation of the SMA is possible, the relative pointing between the high and low frequency receivers can be measured simultaneously by making azimuth and elevation scans across a planet and monitoring the receiver outputs using a synchronous detection scheme. In Figure 7 we present azimuth and elevation scans obtained using this method to demonstrate the relative pointing of the three receivers installed in antenna 1. The largest offset between receivers, ~ 6", occurs between the 230 GHz and 650 GHz units in antennas 1 and 4. More typically we achieve pointing offsets of order 3" – 4" and efficient simultaneous dual frequency operation is possible.  In practice, since the majority of the radio pointing data is obtained using the 230 GHz receivers, we generally use the 230 GHz radio pointing model with the corresponding feed offset when observing with the 650 or 345 GHz receivers.  In order to proceed with dual polarization measurements we have designed and implemented a switching scheme incorporating rotating quarter wave plates for operation at 342 GHz using a single receiver in each antenna.  Eventually dual polarization will be possible in the frequency range 325 – 350 GHz using the orthogonally polarized 345 and 400 GHz receivers.  In this case, assuming co-alignment similar to that of the existing receivers, a pointing intermediate between the two receivers would be used, and cross polarized contamination between receivers would be kept to a minimum via the central wire grid beneath M6.

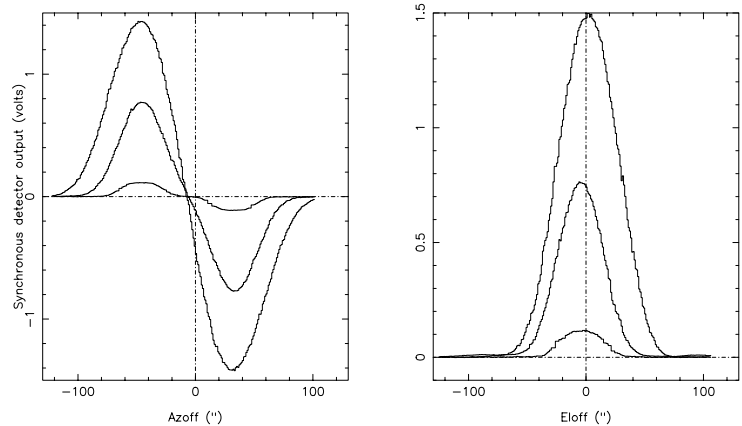

**Figure 7:** Continuum detector output scans in azimuth and elevation across Jupiter showing the relative alignment of the beam positions on the sky of the 3 receivers: 230 GHz (top), 345 GHz (middle), and 650 GHz (bottom) installed in antenna 1.

Referring once more to Figure 6, left panel, we can identify 3 major systems: the optics, the receiver insert, and cryostat infrastructure.  The optical components are used to steer the signal and LO, which is housed on a shelf below M6 but above the individual sets of receiver optics, to a given receiver either by reflection off a rotating wire grid polarizer or a rotating mirror assembly, each of which has 4 positions.  The signal then enters the cryostat through a vacuum window in the receiver insert, and passes through some infrared blocking filters, through a lens cooled to ~ 70 K, and into the mixer via a corrugated feed.  Output from the mixer, at 4 – 6 GHz, is then coupled to a low noise cryogenic amplifier via an isolator equipped with an integrated bias tee.  At this point, the signal passes to the infrastructure components of the cryostat: a 4-way FET switch and band-pass filter, and is further amplified to -30 dBm before exiting the cryostat.

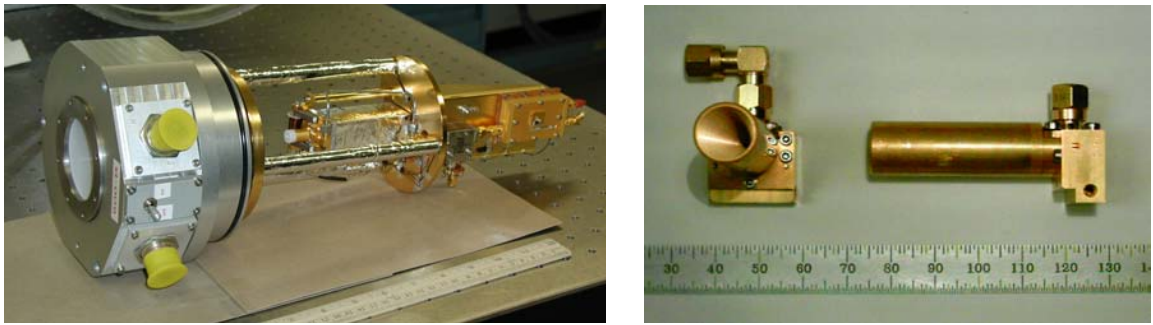

**Figure 8:** Left, photograph of a 650 GHz receiver insert showing, from left to right, the vacuum window, electrical connectors, radial o-ring seal, interface to 70 K radiation shield, mixer feed with Teflon dust cover, magnetic field coil, isolator and 4 – 6 GHz IF preamplifier.  Right, photograph of a 345 GHz SIS mixer block.

In Figure 8 we show a 650 GHz receiver insert in detail, the vacuum window, mixer block with corrugated feed, isolator, and IF amplifier are clearly visible.  The vacuum window is made of Teflon, half a wavelength thick, and the lens, also made of Teflon, is cooled via a link to the cryostat radiation shield at 70 K.  The mixer block and associated IF amplifier are cooled via individual copper heat straps connected to the 3$^{rd}$ stage of a Daikin CG-308SCPR cryo-cooler, which has a cooling capacity in excess of 2 W at 4 K, 2 W at 15 K, and 25 W at 70 K.  In the SMA antennas, the cryo-cooler is operated with its displacer aligned vertically and the cold-surface uppermost.  In this orientation the 3$^{rd}$ stage temperature remains stable to within ± 5 mK on some units, although ± 50 mK is more typical.  Such large temperature changes induce mixer gain fluctuations which result in receiver output fluctuations of up to 1%.  Using a gain stabilization technique developed for the SMA [7] these fluctuations can be reduced to < 0.1% and the receivers can be made sufficiently stable to permit continuum monitoring of the atmosphere for phase correction techniques to be tested.

For simplicity we consider receiver noise to be made up of three major components: that arising due to losses in the receiver optics, mixer noise, and multiplied IF noise.  Generally speaking, losses in the receiver optics can be kept small through the use of reflective optics, and IF amplifiers with a noise level of < 1 K/GHz can be purchased, so the majority of receiver design effort is often placed on mixer design.  In the case of the SMA a complete set of receiver specifications was never developed.  Instead, a goal to provide double side-band receivers to cover the frequency bands specified in Table 5 with the projected noise figures was set.  To this end, we have developed a series of single-ended, fixed-tuned waveguide SIS mixers for the 230, 345, and 650 GHz frequency bands [8,9,10], and a number of laboratory measurements have been made at other frequencies [11].  In all cases, the mixer block is a fixed-tuned, half-height waveguide unit similar to that shown in Figure 8 (right). Referring to the figure, an SIS junction with the required tuning circuit is deposited on a crystalline quartz substrate, which in turn is simply sandwiched between a corrugated feed horn and a copper back-piece containing a waveguide section, machined to the appropriate depth. IF output from the mixer is via an SMA connector mounted on top of the block. This type of mixer mount is extremely simple yet provides a fixed impedance of ~ 25 Ω to the SIS mixer chip over a large input bandwidth.

A sample of receiver noise measurements is given in Figure 9. Referring to the figure, the on-telescope performance appears to be about a factor of 2 worse than that measured in the lab, so it is worth noting the major differences between the measurements. Lab measurements are generally made using a liquid helium test cryostat with the appropriate thickness Teflon vacuum window. This cryostat is equipped with an IF amplifier chain that has a noise temperature of 4 K, measured over a 1 GHz bandwidth, centered at 5 GHz. For the 230 and 345 GHz receiver measurements the lens in front of the mixer feed is mounted on the 4.2 K cold-plate and the infra-red filtering blocking filters are attached to the 77 K radiation shield. In the case of the 650 GHz measurements, the signal is coupled to the mixer via reflecting optics, and LO power is injected via a low-loss beam splitter. Receiver noise measurements on the telescopes are made using room temperature and liquid nitrogen cooled loads placed at the image of the mixer feed aperture between M4 and M5, and consequently include noise arising from losses in the optics above the cryostat (see right panels of Figures 4 an 6). At all frequencies, the lenses in front of the mixers are cooled via a connection to the 70 K radiation shield, and are significantly warmer than in the lab test cryostat, and LO power is coupled to the 650 GHz receivers using Martin Puplett diplexers. Finally, all receiver noise measurements are made using IF amplifier chains that have a noise temperature of typically 10 – 12 K averaged over the full 2.5 GHz IF bandwidth. At the present time we are developing a receiver design for the 325 – 425 GHz frequency range to provide increased sensitivity and to enable more efficient dual polarization measurements to proceed. Developments to improve on-telescope performance of existing receivers will occur as time permits.

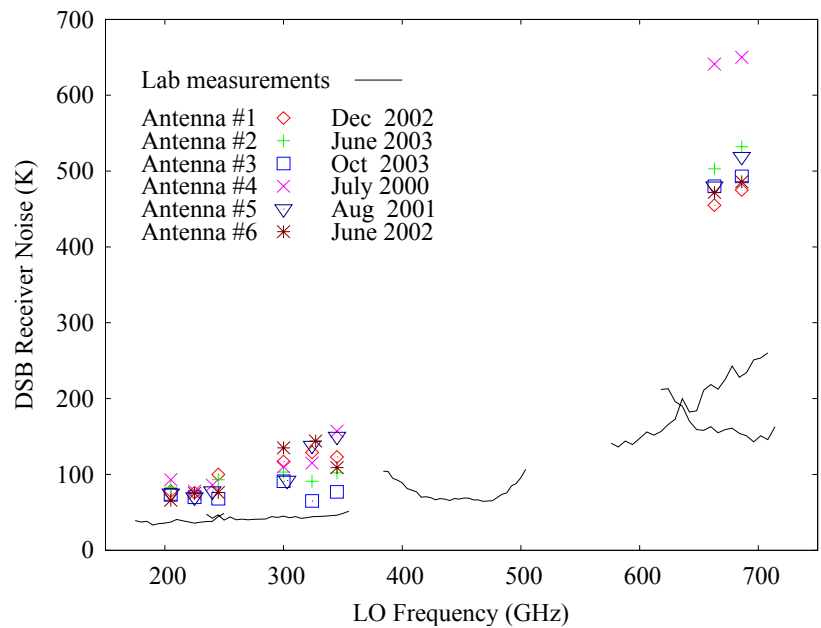

**Figure 9: The measured lab performance of receivers developed for the SMA and the on-telescope performance of all 18 SAO built SMA receivers is plotted as a function of frequency. The discrepancies between lab and on-telescope performance are understood and improvements to receivers deployed the on the SMA antennas will be made as time permits.**

Finally, in Figure 10 we show a picture of the current status of the array, astronomical observations are proceeding in the 230, 345, and 650 GHz bands, and will shortly be open for general observing.

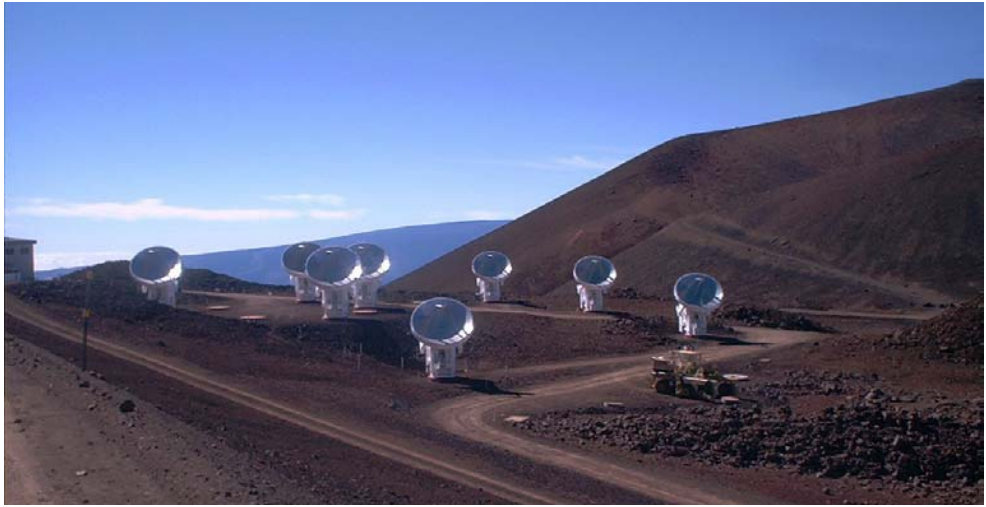

**Figure 10: All 8 antennas of the SMA observing during the SMA dedication, November 22nd 2004.**